# Suppression of Superfluid Density and the Pseudogap State in the Cuprates by Impurities

Unurbat Erdenemunkh,[1] Brian Koopman,[1] Ling Fu,[1] Kamalesh Chatterjee,[2] W. D. Wise,[2] G. D. Gu,[3] E. W. Hudson,[4,2] Michael C. Boyer[1,2*]

[1]Department of Physics, Clark University, Worcester, MA 01610, USA

[2]Department of Physics, Massachusetts Institute of Technology, Cambridge, MA 02139, USA

[3]Condensed Matter Physics and Materials Science Department, Brookhaven National Laboratory, Upton, New York 11973, USA

[4]Department of Physics, Pennsylvania State University, State College, PA 16802

[*]To whom correspondence should be addressed: mboyer@clarku.edu

**We use scanning tunneling microscopy (STM) to study magnetic Fe impurities intentionally doped into the high-temperature superconductor $Bi_2Sr_2Ca_2CuO_{8+\delta}$. Our spectroscopic measurements reveal that Fe impurities introduce low-lying resonances in the density of states at $\Omega_1 \approx$ 4meV and $\Omega_2 \approx$ 15 meV allowing us to determine that, despite having a large magnetic moment, potential scattering of quasiparticles by Fe impurities dominates magnetic scattering. In addition, using high-resolution spatial characterizations of the local density of states near and away from Fe impurities, we detail the spatial extent of impurity affected regions as well as provide a local view of impurity-induced effects on the superconducting and pseudogap states. Our studies of Fe impurities, when combined with a reinterpretation of earlier STM work in the context of a two-gap scenario, allow us to present a unified view of the atomic-scale effects of elemental impurities on the pseudogap and superconducting states in hole-doped cuprates; this may help resolve a previously assumed dichotomy between the effects of magnetic and non-magnetic impurities in these materials.**

Following the idea that "if we know what breaks it, we can figure out what holds it together," many have studied the impact of elemental impurities intentionally doped into high-temperature superconductors (HTS) with the goal of gaining insight into their pairing mechanism. Early doping studies demonstrated that, in contrast to what happens in conventional superconductors, non-magnetic impurities were more detrimental to superconductivity than magnetic ones [1-3] and this generalization appeared confirmed at the atomic-scale by scanning tunneling microscopy (STM) studies [4,5]. Impurities, whether native or intentionally doped into the bulk or at the surface of a superconductor, generate local, low-lying electronic states. These "quasiparticle resonances" have been imaged by STM, which, with its ability to measure spectroscopic maps indicative of the spatially varying local density of states, allows characterization of their energetic and spatial structure.[4-10] In particular, in doped $Bi_2Sr_2Ca_2CuO_{8+\delta}$ (BSCCO) STM measurements in the vicinity of "non-magnetic" Zn impurities find a single impurity resonance near the Fermi energy accompanied by suppressed spectral peaks, which have been interpreted as indicative of the local destruction of superconductivity through strong potential scattering.[4] In contrast, STM measurements near "magnetic" Ni impurities find a split resonance, due to an exchange interaction of the quasiparticle spin with the magnetic moment of the Ni impurity atom, as well as complementary particle-hole quasiparticle states, suggesting the preservation of superconductivity in the impurity affected region.[5] These atomic-scale studies point to potentially marked differences in the local effects of non-magnetic and magnetic impurities on high temperature superconductivity. However, given that Ni impurities in BSCCO have a relatively small effect on $T_C$ compared to other magnetic impurities such as Fe and Co [11,12], comparative STM studies of other magnetic impurities will allow for a more complete understanding of the relative impact of potential and magnetic scattering by impurities on the superconducting and pseudogap states.

To address this we have studied Fe impurities in BSCCO, where, at low concentrations, Fe impurities can reduce $T_C$ at five times the rate of Ni impurities.[11] Similar to Zn and Ni, Fe impurities substitute for Cu atoms in the $CuO_2$ plane. Thermopower measurements indicate that Fe is in the $Fe^{2+}$ ionic state when substituting for $Cu^{2+}$ in the $CuO_2$ plane, as evidenced by no change in sample hole concentration in near optimally-doped samples for Fe doping of less than ~4%.[12] This indicates that the effects Fe impurities have on bulk $T_C$ are not due to changes in carrier (hole) concentration but rather are linked to impurity interactions with quasiparticles in the $CuO_2$ plane. Magnetic susceptibility measurements of Fe impurities (in the $3d^6$, s = 2 state) indicates a ~5.2 $\mu_B$ magnetic moment in $Bi_2Sr_2Ca_2CuO_{8+\delta}$ [12], considerably larger than the ~1.5 $\mu_B$ estimated magnetic moment of Ni.[13]

Our $Bi_{2.1}Sr_{1.9}Ca(Cu_{(1-x)}Fe_x)_2O_{8+\delta}$ samples (x = 0.005) have a $T_C$ of 82 K, reduced by the Fe from the 89 K $T_C$ of their slightly oxygen-overdoped parent (x = 0) material.[14] We cleave our samples in ultrahigh vacuum at 20 K and immediately insert them into our home-built, variable-temperature STM. We initially characterize them by acquiring atomically resolved topographies and high energy resolution spectroscopic maps over extended spatial regions. Similar to previous STM work on BSCCO [15-17], we find spectra with a uniform spectral kink at ~24 mV inside a varying larger gap (**Fig. 1a**). Following our earlier work [18] in the two-gap scenario [19], we associate the 24 mV kink with the superconducting gap, $\Delta_{SC}$, and the larger gap as the inhomogeneous pseudogap (**Fig. 1b**).

In addition to these ubiquitous gap related features, we also find an intragap peak at ~+16 mV associated with Fe impurities, which we can use to map out their locations (**Fig. 1c**). Fe impurity resonances display a 4-fold symmetric spatial structure similar to previously studied Zn and Ni impurities (**Fig. 2a**). More specifically, the '+-shaped' spatial pattern and spatial extent seen at +16 mV for Fe impurities is very similar to the patterns seen near 0 mV for Zn impurities [4] and at positive biases for Ni impurities [5]. Confirming that the +16 mV resonances are indeed associated with Fe impurities, we find 81 such resonances at +16 mV over a 500 Å-square region, representing an impurity doping of ~0.25 %, consistent with the nominal Fe doping of 0.5%. It is not surprising to see variations from nominal doping due to solubility issues; such variations were seen in STM measurements of Zn impurities.

However, rather than being centered on surface Bi atoms, and hence close to subsurface Cu atoms, as is the case for Zn and Ni impurity resonances [4,5], we find Fe resonances to be centered 1.0 - 1.5 Å away from the expected site (**Fig. 2c**). This spatial shift cannot be explained by the presence of a supermodulation, which shifts Cu sites in the $CuO_2$ plane relative to Bi sites in the Bi-O layer by at most 0.4 Å along the a-axis.[20] Rather, this seemingly surprising shift can be explained by the Fe impurity's differing coordination number on substitution. $Cu^{2+}$ ions in BSCCO have a coordination number of five due to their square-pyramidal oxygen binding geometry. When substituting for $Cu^{2+}$, both $Zn^{2+}$ and $Ni^{2+}$ ions can also have a coordination number of five, but such a coordination number is unfavorable for $Fe^{2+}$ ions.[21,22] Since $Fe^{2+}$ favors a different coordination number (four or six), its introduction into BSCCO will tend to induce local lattice distortions as evidenced by the observed position shift of Fe in our data. A similar explanation was used to explain x-ray measurements detecting increases in *a-b* lattice parameters and a decrease in the *c*-lattice parameter when substituting $Co^{3+}$ for $Cu^{2+}$; $Co^{3+}$ favors a coordination number of six which, in turn, induces local lattice distortions.[23] Furthermore, a comparison of *c*-lattice parameters with doping of Ni, Co, and Fe impurities in BSCCO shows a systematic decrease in the *c*-lattice parameter with Co and Fe doping, but virtually no change with Ni doping,[11] consistent with this explanation and consistent with the differing locations of Ni and Fe impurities as detected by STM. Because of this spatial shift we utilize a naming scheme which is not tied to Cu sites when describing the spatial evolution of the resonance (**Fig. 2d**). We note that despite this spatial shift, the observed DOS modulations are near-commensurate with the underlying lattice, as has been previously seen in STM studies of other impurities.[4,5,9,10]

Similar to sample wide averages, "background" spectra taken directly outside of the Fe impurity location of Fig. 2 show kinks at $\varDelta_{SC} \sim 24$ mV and gap edges at $\varDelta_{PG} \sim 33$ mV (**Fig. 3**). Spectra acquired at the north, center, and south regions of the impurity-affected region show an additional impurity induced resonance at +16 mV (**Fig. 3a**). Similar to what is observed near Ni impurities, the spectral gap peaks (defining $\Delta_{PG}$) in the Fe impurity-affected region are unchanged from that of the local background.[5] However, these Fe resonances differ from the Ni resonances in two important ways.

First of all, Ni resonances display a spatially complementary particle-hole symmetry [5]. That is, peaks at $+\Omega$ at the center of the resonance appear to move to $-\Omega$ in the "nearest neighbor" locations (C2NW, C2NE, C2SW, and C2SE in our notation). The existence of spatially integrated particle-hole symmetry suggests local preservation of superconductivity [24]. In our measurements we find no obvious complementary impurity structure at -16 mV (**Fig. 2b**) nor clear evidence for complementary peaks at negative bias in C2NW or C2SE spectra (**Fig. 3b**) or in spectra spatially averaged over the impurity region (**Fig. 3c**). Compared to Ni impurities, this lack of obvious complementary behavior indicates, at minimum, a partial suppression of superconductivity by Fe impurities. Secondly, and perhaps more surprisingly, our Fe impurity spectra show no clear evidence for magnetic splitting of the resonance peak. Such splitting, resulting in two same bias peaks, is both theoretically expected for magnetic impurities [25] and observed in Ni impurities [5].

Looks, however, can be deceiving. In order to enhance subtle spectral features that change spatially we normalize (divide) our spectra by the local background:

$$G_N(E) = \frac{G_{Impurity}(E)}{G_{Background}(E)}$$

This normalization reveals an additional impurity resonance at +4 mV (**Fig. 4a**), evidence for the expected interaction of quasiparticle spin with the magnetic moment of the Fe impurity atom. Similar results can be obtained using a subtractive normalization scheme (see supplemental materials).

To quantify the scattering strength of Fe impurities, we employ the model for quasiparticle potential and magnetic scattering by impurities developed by Salkola et al.[25] which was previously used to quantify STM-studied Zn and Ni resonances [4,5].

$$\frac{\Omega_{1,2}}{\Delta_0} = \frac{-1}{2N_F(U \pm W)ln|8N_F(U \pm W)|}$$

Here $\Omega_{1,2}$ represents the impurity induced resonance energies, $\Delta_0$ represents the spectral gap size, $N_F$ represents the normal-state density states at the Fermi energy, $U$ represents the coulomb interaction between a quasiparticle and an impurity atom, and $W$ represents the magnetic interaction between a quasiparticle and the magnetic moment of the impurity atom. In the two-gap scenario, there is some ambiguity as to whether the superconducting or pseudogap gap value should be used for $\Delta_0$. While the presence of complementary particle- and hole-impurity resonances are set by the presence of the superconducting state [26,27], previous temperature-dependent STM measurements of native impurities in related $Bi_2Sr_2CuO_{6+x}$ clearly indicate that the energy and spatial distribution of the main impurity resonances are set by the pseudogap state, not the superconducting state.[10] For this reason we set $\Delta_0 = \Delta_{PG}$.

**Table 1** summarizes our calculated potential and magnetic scattering values for Fe impurities in comparison to previously studied Ni and Zn impurities. We note that the previous Zn and Ni calculations [4,5] also set $\Delta_0$ equal to the large spectral gap, now known to be the pseudogap, which allows for direct comparison of our Fe impurity calculations to those of Zn and Ni. Our calculations show that, despite a large magnetic moment, potential scattering dominates magnetic scattering of quasiparticles by Fe impurities, as was similarly determined to be the case for Ni impurities. In addition, in comparison to Ni impurities, we find Fe impurities to be both stronger potential (1.7 times larger) and magnetic scatterers (3.4 times larger). The larger magnetic scattering strength of Fe impurities compared to Ni impurities is consistent with its ~3.5 larger magnetic moment.

These normalized spectra not only reveal the expected magnetic splitting of the resonance, but also show at least weak spatially integrated particle-hole symmetry. That is, when we normalize a spectrum, spatially-averaged across the impurity-affected region, to the local background, we find a complementary peak at -15 mV in addition to the strong resonance at +15 mV as well as evidence for complementary lower energy peaks near ±3 mV (**Fig. 4b**). The presence of these complementary peaks indicates at least partially-preserved particle-hole symmetry in the impurity affected region indicating the presence of superconductivity in the immediate vicinity of the impurity.

The spatially averaged normalized impurity spectrum shows two additional important features. First, dips near ± 33 mV indicate a suppression of the pseudogap state in the impurity-affected region; the spectral peak heights of the pseudogap state are diminished in the impurity affected region but the peak locations are unchanged. This local effect of impurities on the pseudogap state is consistent with Raman measurements on impurity doped $Bi_2Sr_2Ca_2CuO_{8+\delta}$ and $YBa_2Cu_3O_{7-\delta}$; the $B_{1g}$ (antinodal) response peak intensity is diminished but energy location is unchanged with impurity doping.[14,28,29] Second, while complementary particle-hole resonance peaks indicate the presence of the superconducting state in the impurity-affected region, the clear dip near -24 mV indicates that the coherence peaks associated with superconductivity are partially suppressed. Because the amplitude of coherence peaks is linked to superfluid density [30-33], the observed spectral suppression at $\Delta_{SC}$ is consistent with a local suppression of superfluid density near Fe impurities. This would explain $T_C$ suppression with Fe impurity doping. Recently, Parham et al. utilized a tomographic density of states technique to analyze angle resolved photoemission spectroscopy data taken on Fe-doped $Bi_2Sr_2Ca_2CuO_{8+\delta}$ and concluded that pairing strength is unaffected near Fe impurities but pair lifetime is reduced [34], also consistent with our atomically resolved spectroscopic STM measurements.

Our studies detailing how Fe impurities affect the local superconducting and pseudogap states in BSCCO allow us to reinterpret previous STM studies of intentionally doped and native impurities and obtain a more complete understanding of how magnetic and non-magnetic impurities interact with these two states. Previous STM measurements on magnetic Ni impurities indicate virtually no local effect on the superconducting state [5] but evidence for partial suppression of the pseudogap state (see supplemental materials section for supporting analysis). In contrast, our studies on magnetic Fe impurities show these impurities lead to a local partial suppression of both. In fact, the local effects of Fe impurities on these states are more comparable to the local effects of non-magnetic Zn impurities, when interpreted in the two gap scenario. Thus, in contrast to previous conclusions, we suggest there is no clear dichotomy between the effects of magnetic and non-magnetic impurities in BSCCO but rather that an impurity's ability to suppress superconductivity, in particular superfluid density, is dominated by its potential scattering strength.

The study of Zn- $Bi_2Sr_2CaCu_2O_{8+\delta}$ by Pan et al. [4] shows background spectra with spectral kinks at ~25 mV (the superconducting gap) in addition to the spectral peaks at ~40 mV (the pseudogap). While Zn impurities destroy the latter, the former survive, suggesting strong suppression of the pseudogap by Zn impurities but locally preserved superconducting order (though reduced superfluid density) with minimal, if any, change in pairing strength. These effects are analogous to what we observe for Fe impurities. We note that this interpretation is consistent with bulk studies indicating both a stronger suppression of the pseudogap state [35-40] and superfluid density [2,41] by Zn impurities than by Ni impurities in the related high-temperature superconductor $YBa_2Cu_3O_{7-\delta}$. In particular, our conclusions are compatible with the "Swiss-Cheese" model previously employed to explain muon spin relaxation rate measurements as a function of Zn doping.[41] In this model, it is assumed that the superfluid density is zero around Zn impurities on the length scale of the in-plane coherence length. Our studies suggest that the superfluid density, while diminished, is not zero in the vicinity of Zn impurities, which explains why the Swiss-Cheese model appears to slightly underestimate the measured relaxation rate (proportional to superfluid density) in Zn-doped $YBa_2Cu_3O_{7-x}$.[41]

This interpretation is also consistent with previous STM studies of native [7,9] and surface impurities [8] in BSCCO which, in addition to the presence of a low-lying impurity resonance, show a narrowing of spectral peaks resulting in a smaller spectral gap. In light of our work, the broad spectral peaks bounding "the gap" in background spectra of these studies actually represent the sum of peaks due to the superconducting and pseudogap states. Near impurities the higher-energy pseudogap peaks are significantly suppressed leaving behind sharper, lower energy, superconducting coherence peaks. Similar to Fe, Ni, and Zn impurities, these native and surface impurities suppress the pseudogap while locally preserving superconductivity, albeit with perhaps some loss of superfluid density.

In each of the hole-doped cuprates $Bi_2Sr_2Ca_2CuO_{8+\delta}$, $YBa_2Cu_3O_{7-\delta}$, and $La_{2-x}Sr_xCuO_4$, bulk measurements of single crystals find that, as long as impurity doping levels are low enough that off Cu-site doping and solubility limits can be ignored, Zn impurities suppress superconductivity (reduce $T_C$) to a greater extent than Ni impurities.[3,42,43] While Zn impurities have a larger effect on $T_C$ than Fe impurities in BSCCO [11,12,42,44,45], in $La_{2-x}Sr_xCuO_4$, the opposite is true [43]. Resistivity [46] and angle-resolved photoemission [47-49] measurements on Zn impurities in $La_{2-x}Sr_xCuO_4$ on near-optimally and overdoped samples indicate, similar, to $Bi_2Sr_2Ca_2CuO_{8+\delta}$ and $YBa_2Cu_3O_{7-\delta}$, that Zn impurities are strong potential scatterers near the unitary limit which can explain $T_C$ suppression with Zn doping consistent with the Swiss Cheese model [41]. Xiao et al. initially attributed the stronger suppression of $T_C$ by Fe impurities in $La_{2-x}Sr_xCuO_4$ to magnetic pair breaking due to Fe's larger retained magnetic moment.[43] However, other studies of Ni, Co, and Fe impurities in $La_{2-x}Sr_xCuO_4$ conclude that their retained magnetic moments cannot explain their effect on superconductivity and resulting $T_C$ suppression.[50,51] How these traditionally magnetic impurities disrupt superconductivity and suppress $T_C$ in $La_{2-x}Sr_xCuO_4$ remains an open question with suggested possibilities including potential scattering by impurities [52,53], carrier localization [50,51], and pair breaking away from nodal regions of the Fermi surface [54]. In addition, the stronger suppression of $T_C$ by Fe than by Zn impurities in $La_{2-x}Sr_xCuO_4$ may reflect the complexities individual to each cuprate family as well as differences between single and multiple layer cuprates. Whereas, in conventional superconductors, magnetic scattering is the dominant mechanism by which impurities suppress superconductivity, our studies suggest it plays only a secondary role to potential scattering in the cuprates. Understanding why the magnetic component of an impurity appears to play such a minor role may ultimately help us to understand what binds together Cooper pairs in these enigmatic superconductors.

**Acknowledgements:** We thank Bill Atkinson and Kyle Shen for useful conversations. The authors thank J. C. Davis for access to Zn and Ni doped Bi-2212 data. This work is supported by NSF grant DMR-1341286 and Clark University (university and physics department research student support). The work at the BNL was supported by DOE, Office of Science under Contract No. DE-AC02-98CH10886.

|         | Zn        | Fe      | Ni       |
|---------|-----------|---------|----------|
| $\Delta_0$ | 44 meV    | 33 meV  | 28 meV   |
| $\Omega_1$ | -1.5 meV  | 4 meV   | 9.2 meV  |
| $\Omega_2$ | ---       | 15 meV  | 18.6 meV |
| $N_F U$    | 4.18      | -1.14   | -0.67    |
| $N_F W$    | ---       | 0.48    | 0.14     |

**Table 1 – Potential and magnetic impurity scattering comparison.** A compilation of gap size, $\Delta_0$, impurity peak locations, $\Omega_1$ and $\Omega_2$, and potential ($N_F U$) and magnetic scattering ($N_F W$) by Zn, Fe, and Ni impurities analyzed in the model developed by Salkola et al.[25] and organized by decreasing potential scattering left to right. Zn and Ni data are taken from [4] and [5] respectively. $\Delta_0$ for Fe and Ni represent local gap values whereas $\Delta_0$ for Zn represents an average gap value over an extended region over the sample.

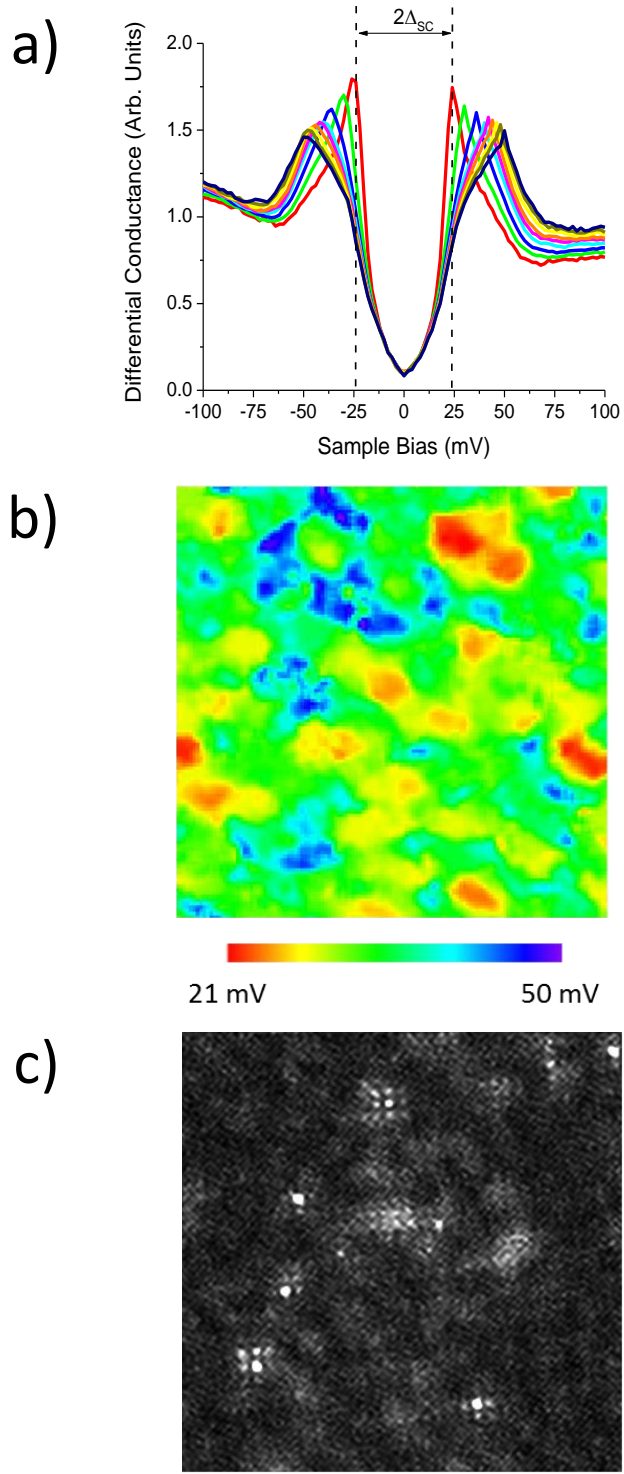

**Figure 1 – Gap inhomogeneity and Fe impurity identification. a)** Spectra that have been group averaged based on $\Delta_{PG}$ from the region in (b). In the smallest gapped spectrum, $\Delta_{PG} \approx \Delta_{SC} \approx 24$ mV. Kinks at 24 mV in the larger gap spectra are associated with $\Delta_{SC}$. **b)** Gap map over a 200 Å square region showing standard inhomogeneity of $\Delta_{PG}$. $\Delta_{PG} = 33 \pm 5$ mV. **c)** +16 mV differential conductance layer taken over a 200 Å region. Several Fe impurities are evident.

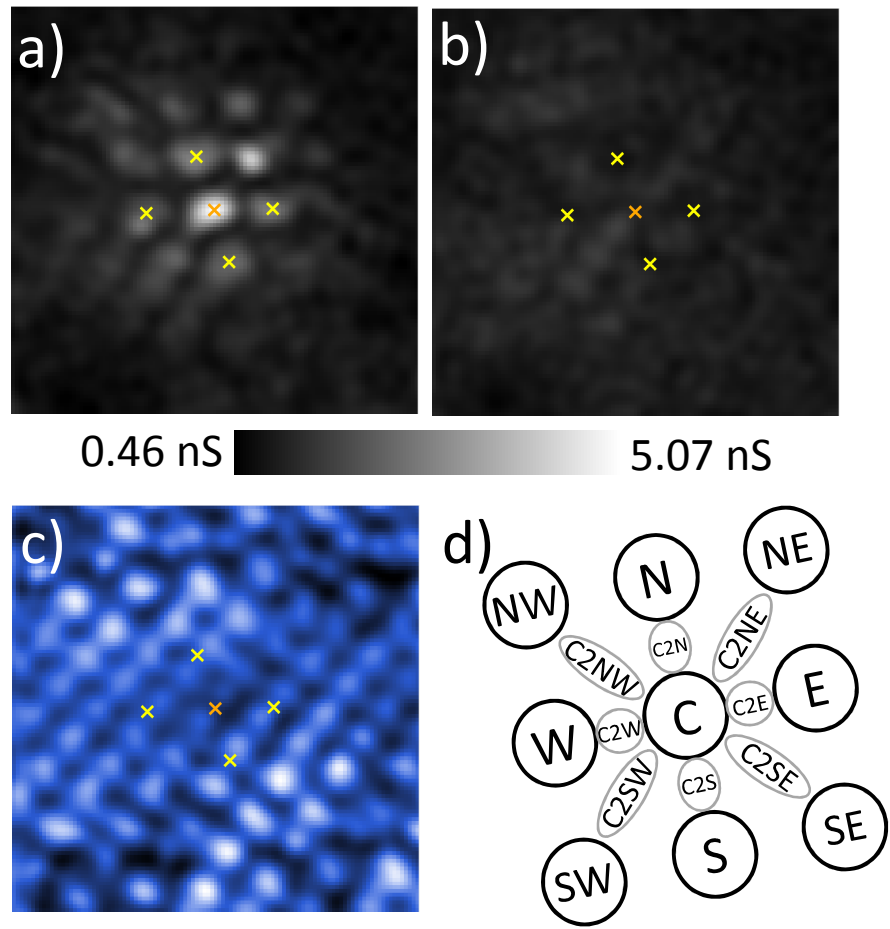

**Figure 2 – Topographic and spectroscopic images around Fe impurity.** 33 Å images taken concurrently around a single Fe impurity. **a)** +16 mV differential conductance map showing the 4-fold symmetrical spatial structure and extent (~25 Å) of the impurity-affected region. An orange 'x' marks the impurity center. Four yellow 'x' marks indicate the locations of the nearest impurity bright spots to the impurity center (N, E, S, and W locations of (d)). These 'x' marks are superimposed on the subsequent images taken over the same location. **b)** -16 mV map of the same location. Unlike previous STM studies of Ni impurities, the Fe impurity affected region does not appear to have complementary particle-hole components. A scale bar in nanoSiemens (nS) has been added below (a) and (b) to emphasize that the color scales for the two map layers are the same. **c)** Topography showing the visible Bi atoms in the Bi-O layer. The impurity center here is shifted ~1.5 Å from the expected Bi atom location. The spatial shifts for five additional Fe impurities are detailed in supplementary materials. **d)** The naming scheme we will use to identify individual features in the impurity affected region. 'C' represents the impurity center as identified by the +16 mV layer. N, E, S, W are the nearest neighbor bright spots in the spectral map layer. NW, NE, SE, SW are the next nearest neighbor bright spots. While the observed impurity pattern in the +16 mV layer for each Fe impurity is the same, we note that there is no universal direction for the impurity center shift relative to the lattice for each Fe impurity. While for consistency we focus on one Fe impurity in Figures 2-4 of our manuscript, we have studied multiple Fe impurities with multiple tips on multiple samples. Our measurements on other Fe impurities are consistent with what we present here.

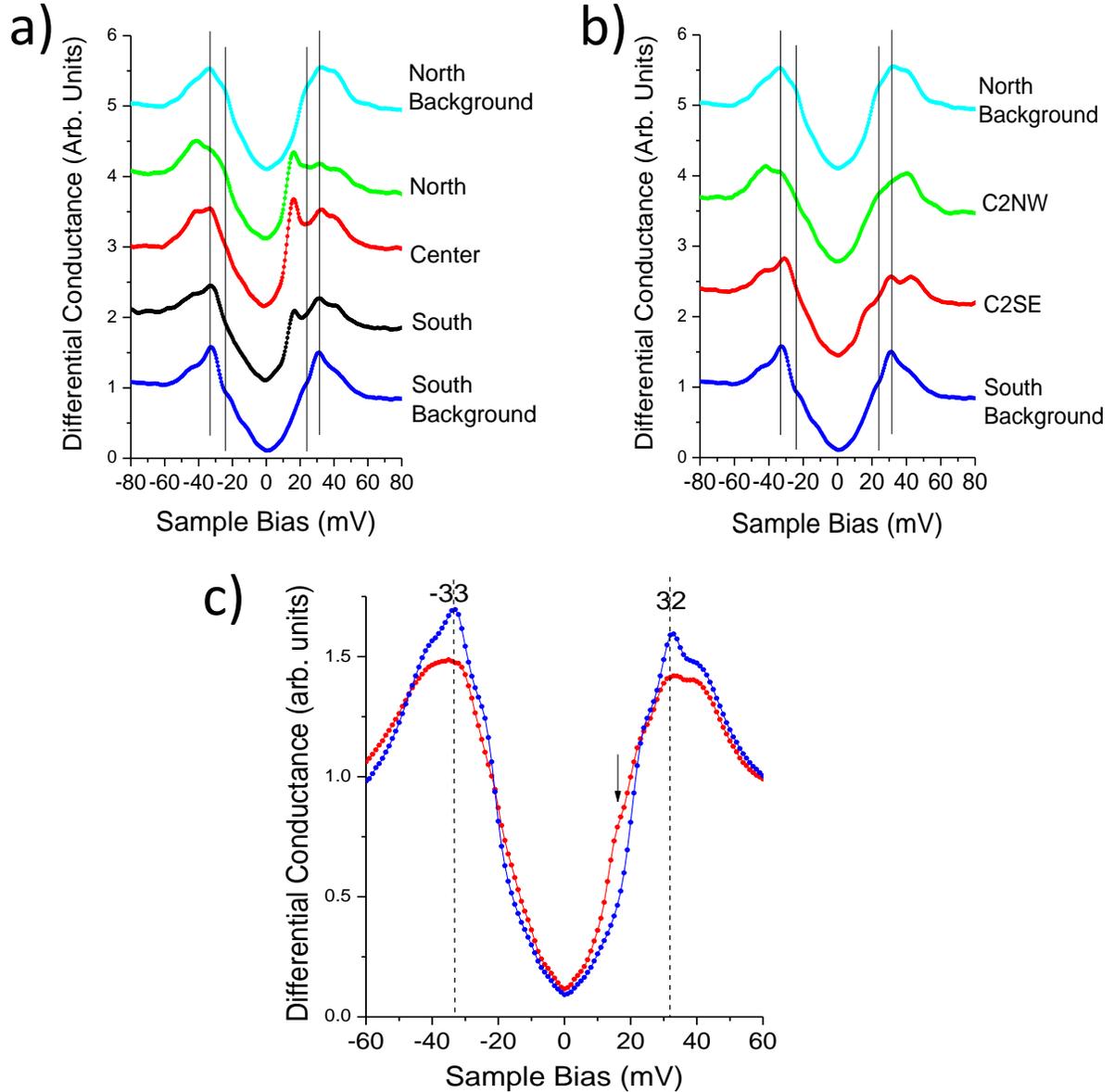

**Figure 3 – Fe impurity and background spectra. a)** "North Background" and "South Background" spectra are taken directly outside of the impurity affected region of Fig. 2 (~15 Å from the impurity center) and show a $\Delta_{SC} \approx 24$ mV kink and peaks indicating $\Delta_{PG} \approx 33$ mV. In addition, the north, center, and south regions of the impurity affected region show a single impurity induced resonance at ~+16 mV. **b)** The C2NW and C2SE spectra show similar gap-defining peaks but no clear sub-gap resonances. **c)** An overlay of a spatially averaged spectrum taken from within the impurity region (red curve) with the average local background spectrum (blue curve). Evidence for the impurity resonance near +16 mV is clearly seen in the red curve. Both the spatially averaged impurity spectrum and the local background spectrum originate from the same 40 Å square spectral map (a 33 Å square region of which is seen in Figure 2). The "local background" spectrum represents an average of spectra from the map which are 15 Å or further away from the impurity center (but still within the 40 Å region).

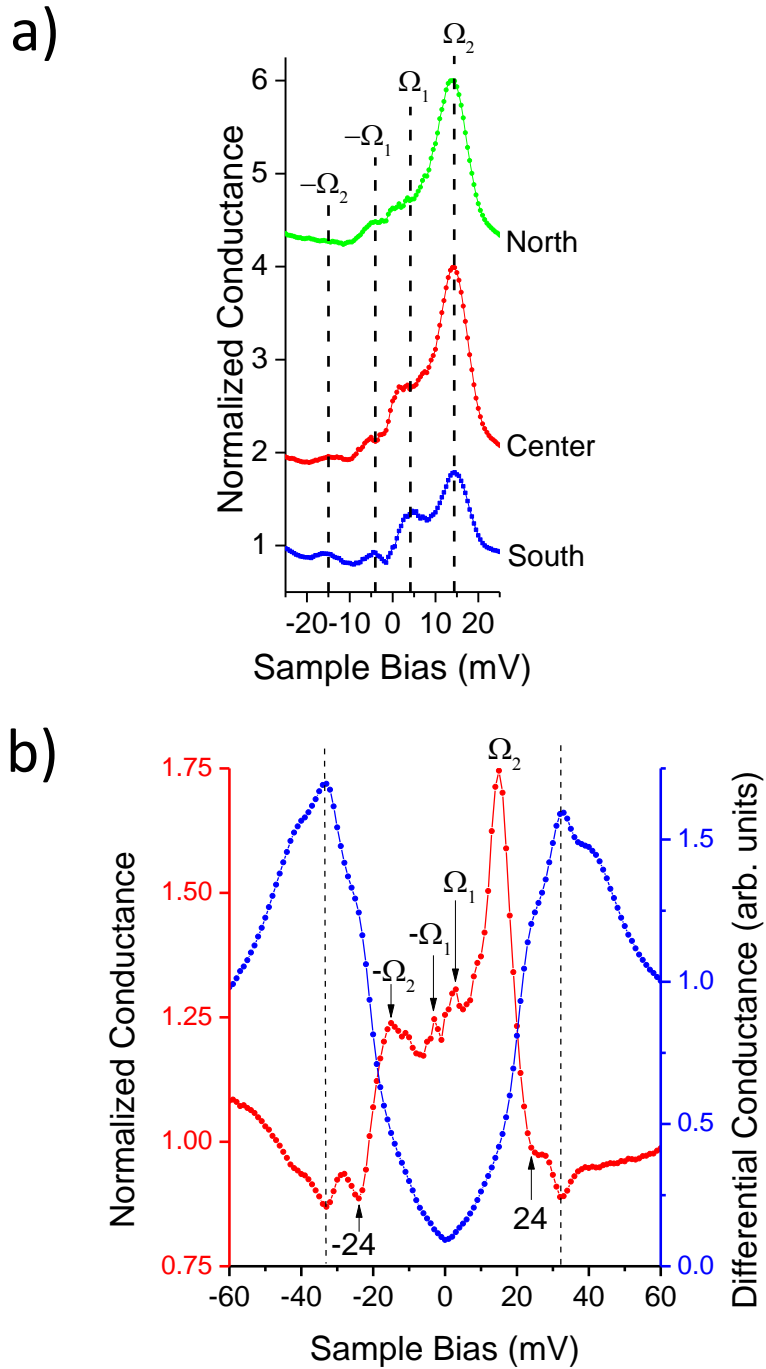

**Figure 4 – Normalized Fe impurity spectra. a)** High-resolution north, center, and south impurity spectra normalized to the local background show evidence for two peaks at positive bias: $\Omega_1 \approx 4$ mV and $\Omega_2 \approx 15$ mV. **b)** Red curve: a spatially averaged spectrum taken from within the impurity region normalized to the average local background spectrum. In addition to the strong +15 mV peak, there appears a distinct complimentary -15 mV peak in the normalized spectrum as well as peaks at ±3 mV. Particle-hole symmetry is partially preserved in the locale of the Fe impurity. The local background spectrum (blue curve) is superimposed on the normalized spectrum for reference.

## Supplementary Material

## Fe Impurity Spatial Shift

**Figure S1** shows high-resolution STM topographic images acquired around five additional Fe impurities (to the one appearing in the manuscript). The topographic images were acquired as part of high-resolution spectral maps. The +16 mV differential conductance layer of the spectral maps were used to identify the location of the Fe impurity centers (indicated by superimposed yellow circles) relative to the Bi atoms appearing in the topography.

In each case, a Fe impurity is shifted relative to the nearest Bi atom indicating that the Fe impurity is shifted from the expected Cu site. Such a shift in the impurity location has not been detected for other impurities (Zn and Ni) substituting for Cu in BSCCO. The shift in Fe impurity location can be linked to $Fe^{2+}$ favoring a coordination number of four or six substituting for a $Cu^{2+}$ ion with a coordination number of five which leads to a local lattice distortion. Furthermore, we note that the shifts of the six Fe impurities studied are not in a uniform direction. These variations may be explained as follows: the Fe-induced lattice distortion is further affected by a) structural shifts due to the supermodulation and b) a varying Coulomb potential due to chemical disorder.

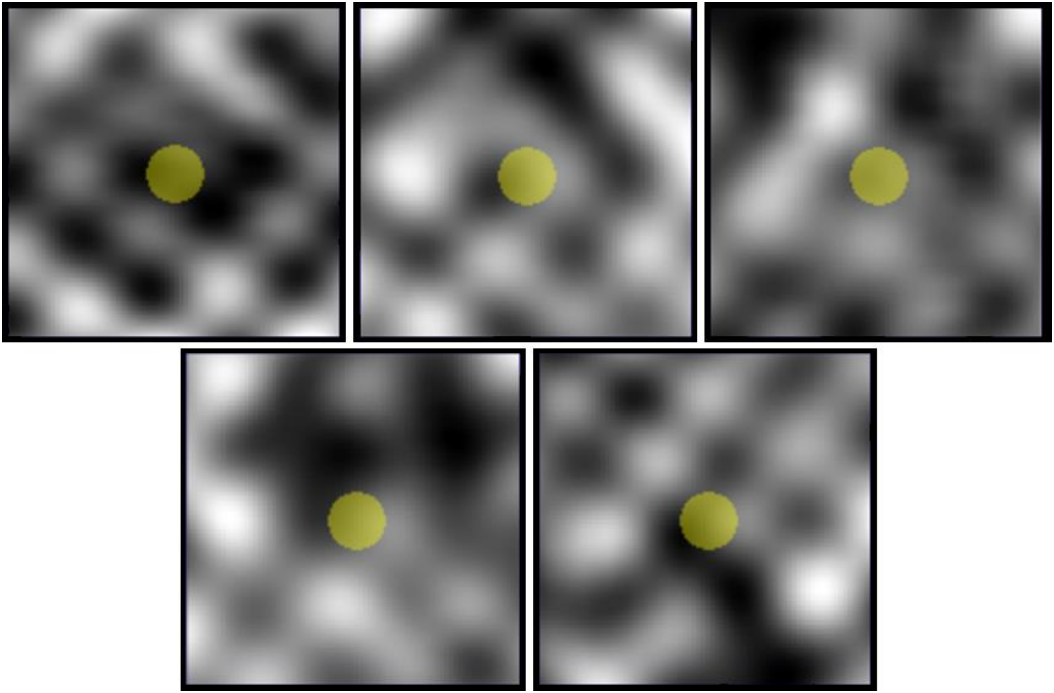

**Figure S1**: **Spatial shift of Fe impurities.** Above are five topographic images taken around Fe impurity locations. The Bi atoms in the topography are seen in white/gray. The Fe impurities are centered in each of the 12 Å square topographies. The +16 mV spectral map layer of spectral maps were used to identify the Fe impurity location. The Fe impurity locations are marked by yellow circles superimposed on the topography. In each case the Fe impurity is shifted from its expected location; they are shifted from directly below a Bi atom and hence shifted from a Cu site.

We reanalyzed previously acquired spectroscopic maps taken on Zn and Ni doped BSCCO utilizing the same spectral normalization scheme ($G_N(E) = \frac{G_{Impurity}(E)}{G_{Background}(E)}$) employed in our analysis of the Fe doped samples.

**Zinc Impurity Analysis**

**Figure S2a** shows spectra which have been group averaged based on spectral gap size across a 500 Å region of the Zn-doped BSCCO sample. The spectra show a wide range of pseudogap gap sizes; $\Delta_{PG} \approx 42 \pm 7$ mV. We utilize the smallest gapped spectrum, consistent with the energy kink in the larger gapped spectra, to determine the superconducting gap size; $\Delta_{SC} \approx 26$ mV.

**Figure S2b** shows an overlay of spatially averaged spectrum taken over a Zn impurity affected region (red) and a spatially averaged local background spectrum (blue) taken directly outside the impurity region. The spectral peaks for the impurity spectrum are of smaller widths and at lower energies than for the background spectrum. This is consistent with at least a partial suppression of peaks defining a larger gap (the pseudogap state) leaving the smaller gap (superconductivity) with greater relative weight in their final sum. **Figure S2c** shows the spatially averaged normalized ($G_N(E) = \frac{G_{Impurity}(E)}{G_{Background}(E)}$) Zn impurity spectrum (red). The background spectrum (blue) is overlaid for reference. The normalized impurity spectrum shows peaks at ±26 mV which corresponds to $\Delta_{SC}$.

At some locations in the sample, the spectral peaks defining the gaps of the superconducting and pseudogap states are separated enough in energy that they can easily be distinguished without employing a normalization technique. **Figure S3a** shows a spectral linecut starting 25 Å away from a Zn impurity and ending at the impurity center. The background spectra away from the impurity show clear peaks defining the superconducting gap $\Delta_{SC} = 28$ mV and the pseudogap $\Delta_{PG} = 39$ mV. Furthermore, it is clear that near the Zn impurity, the peaks of the pseudogap are significantly suppressed and that the peaks defining the superconducting gap survive. **Figure S3b** shows an overlay of the Zn impurity spectrum (red) with a background spectrum 25 Å away (blue). The peaks defining the smaller superconducting gap clearly survive directly at the Zn impurity center.

We conclude that superconductivity exists even within the Zn impurity affected region. In addition, the superconducting gap size does not change within the impurity affected region to that away from the impurity region. This provides strong evidence that the pairing strength is unaffected by Zn impurities. Rather, similar to our conclusions regarding Fe impurities, we conclude that Zn impurities act to suppress superconductivity through a local reduction of superfluid density.

The strong suppression of the spectral peaks suppression by Zn impurities reported in Pan et. al. [1], initially interpreted as suppression of the superconducting coherence peaks, is evidence for strong suppression of the pseudogap state.

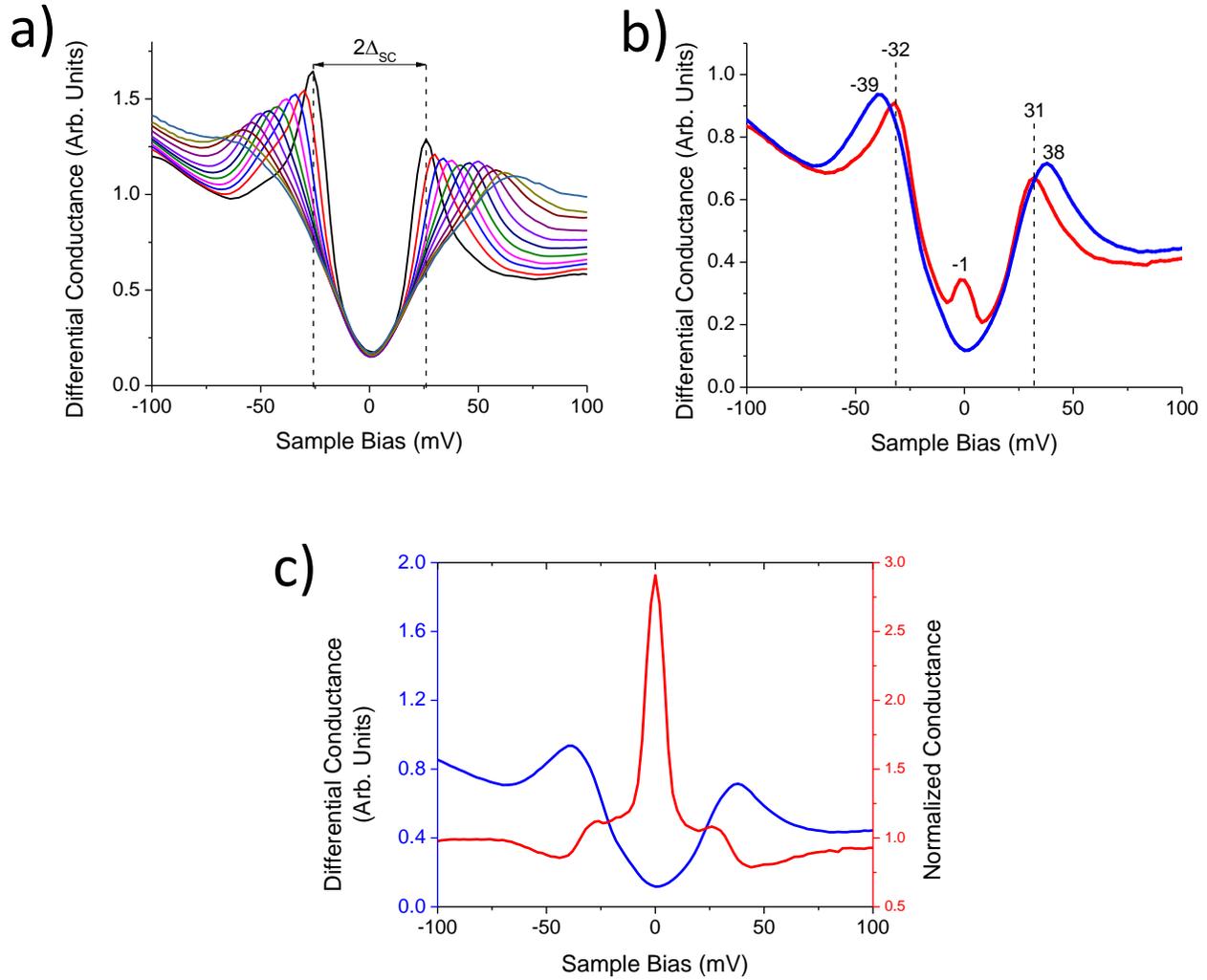

**Figure S2**: **Superconducting gap identification and spatially averaged Zn impurity spectra.** a) Group average spectra from a 500 Å region of Zn-doped $Bi_2Sr_2Ca_2CuO_{8+\delta}$; $\Delta_{PG} \approx 42 \pm 7$mV, $\Delta_{SC} \approx 26$ mV. b) A spatially averaged Zn impurity spectrum (red) is overlaid with a spatially averaged local background spectrum taken directly outside the Zn impurity affected region. c) The average impurity spectrum normalized to the local background reveals peaks at 26 mV in addition to a central impurity peak near 0 mV (red). The background spectrum (blue) is overlaid for comparison.

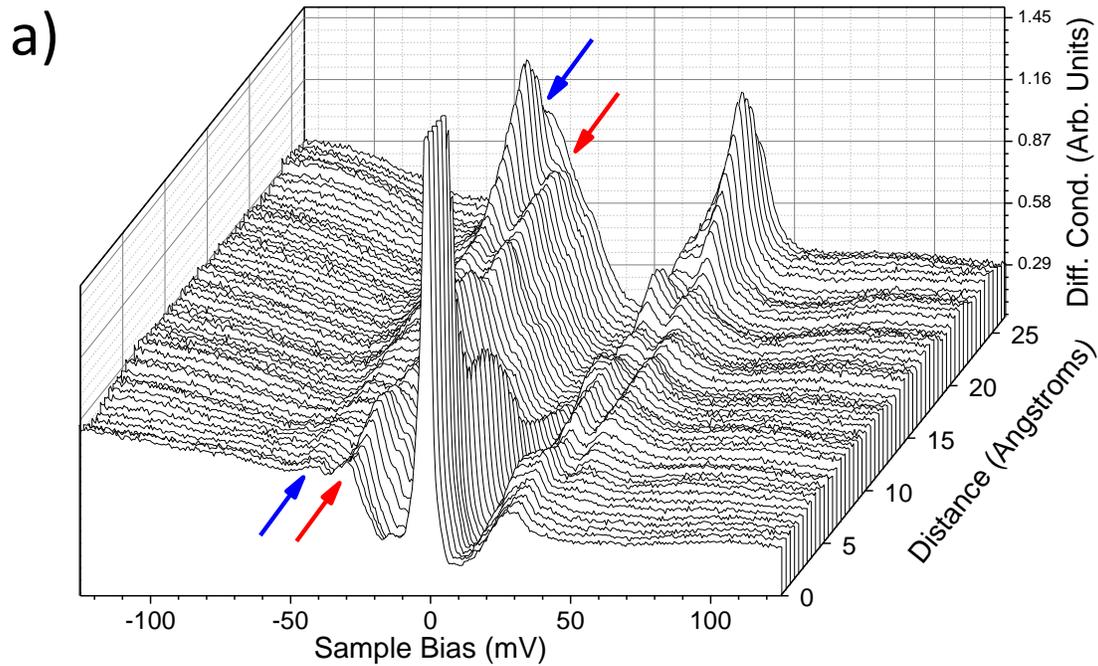

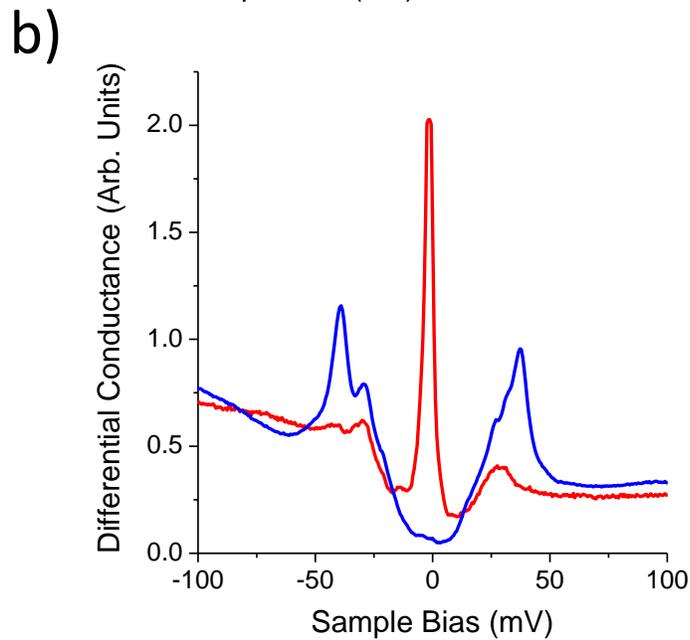

**Figure S3**: **Suppression of pseudogap and superfluid density but preserved superconducting order parameter at Zn impurity site.** a) dI/dV spectra taken starting at a Zn impurity center and extending 25 Å away from the impurity. The peaks defining the gaps of the pseudogap and superconducting spectral gaps are clearly defined. b) Overlay of spectrum taken at the impurity center (red) with a spectrum taken 25 Å away (blue). The peaks defining the inner gap away from the impurity center are clearly still present in the Zn impurity spectrum.

**Nickel Impurity Analysis**

**Figure S4a** shows an overlay of spatially averaged spectra taking over a Ni impurity affected region (red) and a spatially averaged local background spectrum (blue) taken directly outside the impurity region. The spatially averaged Ni impurity spectrum is overall particle-hole symmetric as reported in Hudson et. al. [2] and shows evidence for complementary low energy impurity peaks as "shoulders" near ± 9 mV.

**Figure S4b** shows the spatially averaged normalized ($G_N(E) = \frac{G_{Impurity}(E)}{G_{Background}(E)}$) Ni impurity spectrum (red) with an overlay of the local background spectrum (blue) for comparison. On normalization, all four impurity peaks $\Omega_1 \approx 9$ mV, $\Omega_2 \approx 18$ mV and complementary peaks at negative bias are resolved. The normalized spectrum is overall particle-hole symmetric showing clear evidence for preserved local superconductivity at the Ni impurity location. In addition, the normalized spectrum shows that the outer gap present in the local background is partially suppressed within the impurity affected region. This provides direct evidence that Ni impurities act to locally suppress the pseudogap state in BSCCO. This also explains why Lang et al. previously found Ni impurities in small gap regions of their sample;[3] the Ni impurities act to suppress the larger energy spectral peaks associated with the pseudogap state.

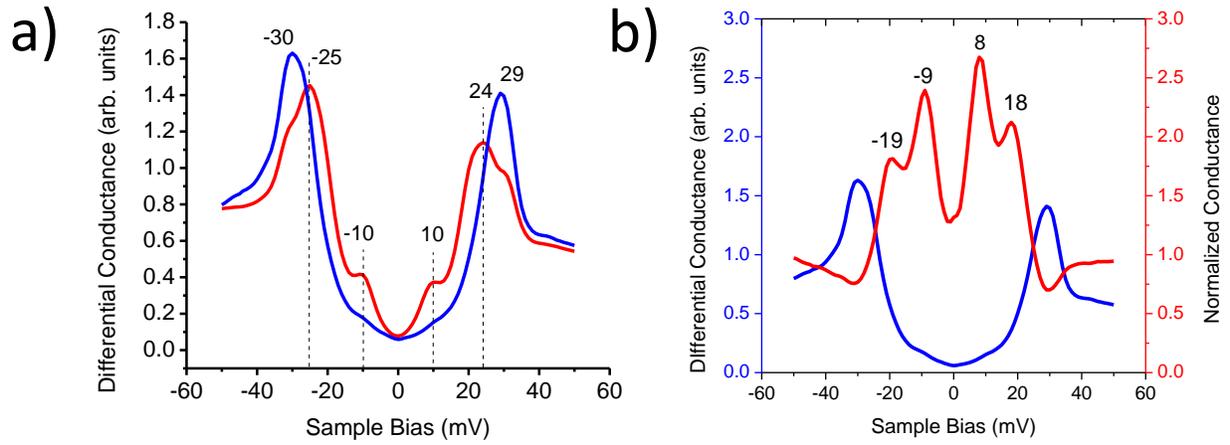

**Figure S4**: **Pseudogap suppression by Ni impurities.** a) A spatially averaged Ni impurity spectrum (red) is overlaid with a spatially averaged local background spectrum (blue) taken directly outside the Ni impurity affected region. b) The average impurity spectrum normalized to the local background reveals impurity peaks at 8 mV, -9 mV, 18 mV, and -19 mV. The background spectrum (blue) is overlaid for comparison.

**Analysis Methodology**

In our manuscript analysis, we employ a normalization scheme to uncover spectral features which are otherwise unseen in the raw data. To determine the specific differences between spectra taken from within an impurity affected region to that directly outside, we divide the impurity affected spectrum by the local background spectrum: $G_N(E) = \frac{G_{Impurity}(E)}{G_{Background}(E)}$

Such a normalization scheme (employing division of STM spectra) has been utilized previously in studying temperature-dependent STM data on the high-temperature superconductor Bi-2201 so as to compare spectra taken below the superconducting transition to those above.[4] While the raw spectra below and above the transition temperature looked nearly identical, this analysis technique revealed important differences: the spectra taken below the superconducting transition temperature had a small gap which was not present in spectra above the transition. We note that the analysis we employ in this manuscript follows a very similar prescription.

However, in employing any analysis technique, we need to be careful that the features produced from such analysis are not artifacts of the normalization scheme. For this reason we also used a second comparative analysis technique in which we subtracted the local background spectrum from an impurity spectrum to elucidate differences between the two. We note that such a subtraction analysis scheme has been used previously to extract differences between STM spectra taken in zero magnetic field in Bi-2201 to those taken in high magnetic fields.[5]

**Figure S5** shows a comparison of the two analysis techniques. The red curve shows the result of dividing the "South" impurity spectrum by the local "South Background" spectrum seen in Figure 3a of the manuscript. This resulting curve is also shown in blue in Figure 4a of the manuscript. Superimposed on the same plot (black curve in Figure S5), is the result when we subtract the local "South Background" from the "South" impurity spectrum.

What is clear is that for either analysis method the resulting features in each are the same. The resulting spectral peaks and dips after either analysis procedure occur at the same energies. We note that division more clearly evinces the spectral peaks; the division process helps to further enhance small differences between curves. For that reason we used that analysis method in our manuscript.

We also note that in employing the normalization scheme, $G_N(E) = \frac{G_{Impurity}(E)}{G_{Background}(E)}$, on Ni impurity data, the impurity peaks are not obvious in the spatially averaged spectrum (Figure S4a) but are obvious after normalization to the local background (Figure S4b). Furthermore, this normalization scheme does not alter the impurity peak energies but rather, after normalization, we find clear peaks are evident at $\Omega \approx \pm 9$ mV and $\Omega \approx \pm 18$ mV in agreement with raw spectra from Ni paper.[2]

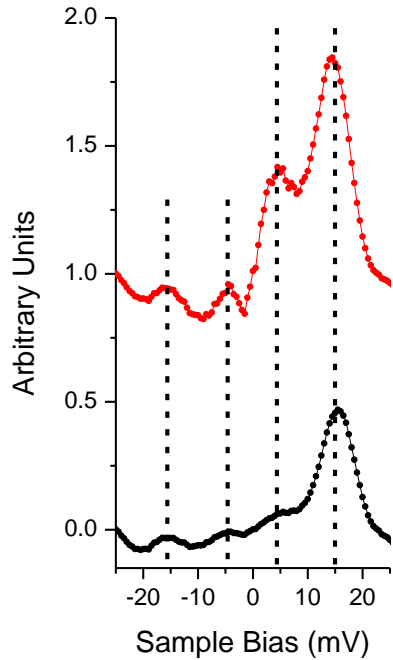

**Figure S5**: **Comparison of analysis procedures.** The red curve represents the "South" impurity spectrum divided by the "South Background" spectrum. Four impurity peaks at $\pm\Omega_1 = \pm 4$ mV and $\pm\Omega_2 = \pm 15$ mV are clearly seen. The black curve shows the "South Background" spectrum subtracted from the "South" impurity spectrum. The same four impurity peaks are seen, but are not as clear. Both analysis techniques show the same impurity features indicating they are robust and not dependent on analysis technique.

Furthermore, if we compare the two analysis processes in comparing the spatially averaged spectrum to the local background, we see the same crucial features using each of the methodologies (**Figure S6**). 1) We see clear peaks at $\Omega = \pm 15$ mV. 2) We see clear suppression at the PG peak energies ($\Delta_{PG} = 33$ mV). 3) We see a clear suppression at -24 mV, the superconducting gap energy, as well as a kink at +24 mV. 4) Finally, the low energy impurity peaks at $\Omega = \pm 3$ mV, obvious through division analysis, are seen as "shoulders" when subtracting the background from the impurity spectra. The features appear in both methodologies, but employing a normalization scheme which involves division enhances the subtle features.

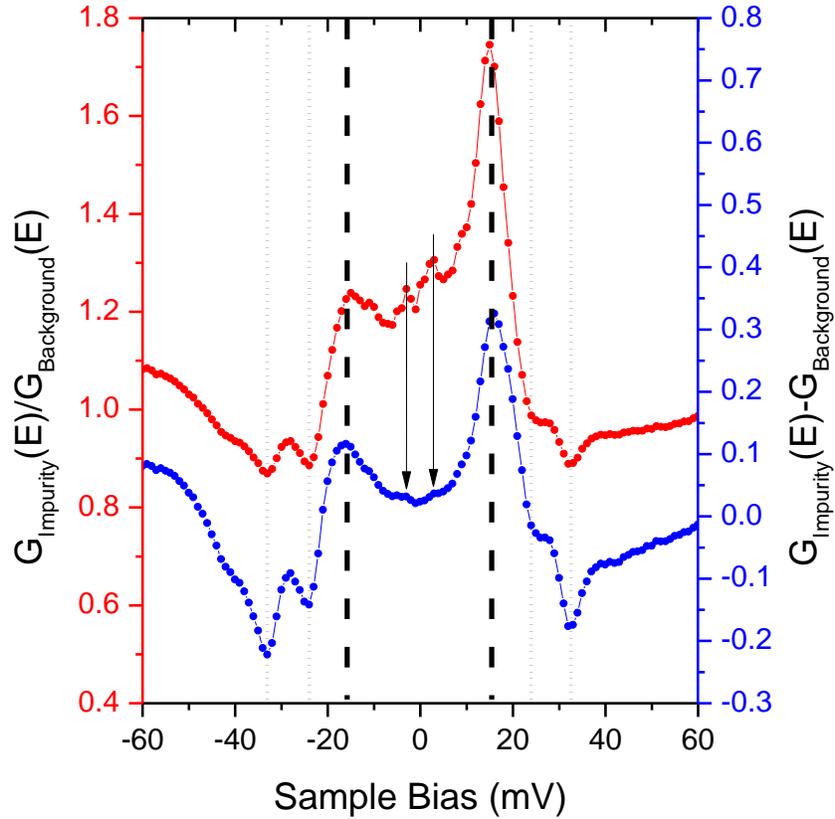

**Figure S6**: **Comparison of analysis procedures for spatially averaged impurity spectrum.** The red curve represents the spatially averaged impurity spectrum seen in Figure 3c divided by the local background from the same figure. The blue curve shows the result of subtracting the local background from the spatially averaged impurity spectrum. The same features are seen in the results of each of the two analysis procedures. Dotted lines and arrows are included as aids to illustrate that the features occur at the same energies.

**Brief Note on Spectral Features:**

The background spectra in Figure 3 of the manuscript show $\Delta_{SC} \approx 24$ mV kink and peaks indicating $\Delta_{PG} = 33$ mV. We note that these spectra also show energy features near 40 mV. The impurity is in a region where $\Delta_{PG}$ changes from ~33 mV to 40 mV and the spectra shown capture features of both pseudogap regions. Given the well-studied spectral gap inhomogeneity in BSCCO which varies on a nanometer length scale, such spectral characteristics are commonly observed in STM spectral measurements on BSCCO and are detailed in [3].

**Final Remarks:**

Our analysis of Fe, Zn, and Ni impurities points to a consistent understanding of the effects of impurities in BSCCO; impurities act to suppress both the superconducting and pseudogap states in BSCCO. These conclusions are consistent with results of studies across impurity doped cuprates by other experimental techniques including ARPES, Raman, μSR, and NMR studies as cited in the manuscript. Our analysis gives specific insight into the atomic-scale effects impurities have on the two states and supports the conclusion that local suppression of superconductivity is linked primarily to the potential scattering strength of the impurity; the effective magnetic moment of a doped impurity ion plays at most a secondary role. This new understanding of impurity effects may help to bring some insight into other seemingly complex cuprate physics including the results of neutron scattering experiments which indicate that Zn and Fe impurities tend to be more effective in establishing static magnetic order in hole-doped cuprates than Ni [6-11]; impurities which are strong potential scatterers are more effective in suppressing the superconducting state which, in turn, has previously been suggested as an important component in allowing for establishment of static spin correlations.